# Multispectral Kerr frequency comb initiated by Faraday ripples

Shu-Wei Huang

**Abstract:**
In a uniform microresonator, the generation of a broadband Kerr frequency comb is triggered by Turing patterns. Here, we study a distinctly different route to initiate the Kerr frequency comb by Faraday ripples. Momentum conservation is ensured by azimuthal modulation of the cavity dispersion. With a good agreement with the theoretical analysis, we demonstrate a multispectral Kerr frequency comb covering telecommunication O, C, L, and 2 μm bands. Comb coherence and absence of a sub-comb offset are confirmed by cw heterodyne beat note and amplitude noise spectra measurements. The device can be used for achieving broadband optical frequency synthesizer and high-capacity coherent telecommunication.

Optical frequency combs, with precisely controlled spectral lines spanning a broad range, have made great impacts on frequency metrology, optical clockwork, precision navigation, and molecular fingerprinting. In addition to the standard implementation based on mode-locked lasers, optical frequency combs based on parametric oscillations in ultrahigh-Q microresonators have become invaluable in applications requiring compact footprint, low power consumption, large comb spacing, and access to nonconventional spectral regions. Recent demonstrations of mid-IR optical frequency microcombs [1–4], low-phase noise photonic oscillators [5, 6], high-repetition rate femtosecond pulse trains [7–11], stabilized optical frequency microcombs [12, 13], and coherent optical telecommunication [14, 15] have revealed the outstanding performance of microresonator-based optical frequency combs, or Kerr frequency combs, and reassured further expansion of already remarkable applications.

For efficient initiation of optical parametric oscillations (OPOs) in a micro-ring resonator, both the energy and momentum conservation laws must be satisfied. In general, the azimuthal mode numbers (N) of the signal and idler are symmetrically located with respect to the pump mode ($\omega_0$), ensuring the momentum conservation, or phase matching condition. On the other hand, satisfying the energy conservation law, $\omega_N + \omega_{-N} - 2\cdot\omega_0 \approx 0$, requires either an intricate balance between the anomalous group velocity dispersion (GVD) and the nonlinear mode pulling effect, or a careful design of the GVD spectral profile, taking into account higher-order dispersions [16, 17]. The generation process is equivalent to spontaneous formation of Turing patterns from a homogeneous background [18], and optical Turing patterns have been studied and utilized for applications including high-capacity communication [14], coherent terahertz wave generation [19], and on-chip optical squeezing [20]. By pumping the micro-ring resonator harder, non-degenerate four-wave mixing (FWM) eventually occurs and fills in the spectral gaps between the first pair of parametric oscillation sidebands, yielding a broadband optical frequency comb [21].

In this letter, we examine and demonstrate another method for initiating OPOs and formation of Kerr frequency combs in ultrahigh-Q microresonators. Here, we consider the case in which the azimuthal mode numbers of the signal and idler are not symmetrically located with respect to the pump mode (Figure 1a). Regardless of the GVD sign and profile, either $\omega_{N+1} + \omega_{-N} - 2\cdot\omega_0 \approx 0$ or $\omega_N + \omega_{-(N+1)} - 2\cdot\omega_0 \approx 0$ hold, i.e., the energy conservation law is automatically fulfilled in this scenario. Figure 1a illustrates the case in which the resonator exhibits only a normal GVD and no higher-order dispersions. Away from the pump mode, the free spectral ranges (FSRs) on the signal and idler sides monotonically decrease and increase, respectively. The mismatch between the signal-pump and pump-idler detuning increases as a quadratic function of the azimuthal mode number, eventually approaching the FSR of the (N+1)[th] signal mode when N equals $\left[\sqrt{\frac{FSR_0}{-D_2}} - 1\right]$, i.e., $\omega_N + \omega_{-N} - 2\cdot\omega_0 = FSR_{N+1}$. Consequently, the energy conservation law of $\omega_{N+1} + \omega_{-N} - 2\cdot\omega_0 \approx 0$ is satisfied. The momentum conservation law, however, is always violated in a homogeneous microresonator and the wavevector mismatch is dk = $2\times\pi/L_{cav}$, where $L_{cav}$ is the cavity length. To fulfill the momentum conservation law, an azimuthal modulation of the cavity parameters has to be introduced to provide an additional wavevector for quasi-phase matching (QPM) [22]. The spontaneous pattern formation mediated by a periodic modulation of system parameters is mathematically equivalent to formation of Faraday ripples, which was first observed in hydrodynamics [23], later expanded to condensed matter physics [24], and recently to fiber nonlinear optics [25–27]. Our use of a $Si_3N_4$ planar microresonator has the advantage of straightforward dispersion management by the design of waveguide geometry (Figure 1b), opening up the possibility of implementing the QPM concept via dispersion modulation along the cavity. The inset of Figure 1b schematically shows the dispersion-modulated single-mode $Si_3N_4$ microresonator, which consists of adiabatically tapered waveguides in the straight region that yield GVD oscillations in the -40–140 fs²/mm range, and uniform single-mode waveguides in the semi-circular region that ensure no excitation of higher-order transverse modes [28]. The microresonator is fabricated in CMOS-compatible processes: first, a 5-µm-thick under-cladding oxide is deposited on a p-type silicon wafer. An 800-nm-thick nitride layer is then deposited using low-pressure chemical vapor deposition, patterned using optimized deep ultraviolet (DUV) lithography, and etched using optimized reactive ion dry etching. Finally, the chip is annealed at 1150 °C for 3 h before the microresonator is over-cladded with a 3-µm-thick oxide layer.

OPOs and comb generation in these cw-laser-pumped nonlinear microresonators can be described by the Lugiato–Lefever equation [27]

$$L_{cav}\frac{\partial}{\partial Z}A = \sqrt{T_c}A_p - \left(\alpha + \frac{T_c}{2} + i\delta\right)A - i\gamma L_{cav}|A|^2 A + \frac{i}{2}\beta_2(Z)L_{cav}\frac{\partial^2 A}{\partial t^2} \quad (1)$$

where $A(Z,t)$ is the envelope function of the intra-cavity electric field, $L_{cav}$ is the cavity length, $T_c$ is the power coupling loss, $\alpha$ is the amplitude attenuation per roundtrip, $\delta$ is the pump-resonance detuning, $\gamma$ is the nonlinear coefficient, and $\beta_2(Z)$ is the GVD profile within the cavity. Of note, the GVD periodicity is guaranteed by the cavity to be $L_{cav}$. Assuming a piecewise constant GVD profile with a duty cycle of 50%, for simplicity, and following the procedures outlined in Ref. [27], we can analytically evaluate the gain of OPOs as a function of the sideband frequency and intra-cavity pump power (Figure 1c). Furthermore, the gain peak sideband frequency can be expressed analytically as

$$\omega_F = \sqrt{\frac{2\left[\left(\frac{\delta}{L_{cav}} - 2\gamma P_{in}\right) \pm \sqrt{(\gamma P_{in})^2 + \left(\frac{\pi}{L_{cav}}\right)^2}\right]}{\overline{\beta_2}}} \quad (2)$$

where $P_{in} = |A_{in}|^2$ is the intra-cavity pump power and $\overline{\beta_2}$ is the path-averaged GVD along the cavity. Under the mean field and good cavity approximations [29], i.e., assuming $\gamma P_{in}, \delta \sim O(\varepsilon)$, Equation 2 can be reduced to

$$\overline{\beta_2}\omega_F^2 = 2\frac{\pi}{L_{cav}} \qquad (3)$$

which explicitly shows how the GVD-induced phase mismatch (left hand side) is compensated by the additional wavevector provided by the GVD modulation along the cavity (right hand side). Such an additional wavevector enables efficient sideband generation even in normal GVD microresonators without any higher-order dispersions, thereby expanding the parametric range in which a Kerr frequency comb is obtained.

Figure 2a shows the path-averaged GVD and third order dispersion (TOD) of our dispersion-modulated single-mode microresonator, simulated using a commercial full-vectorial finite-element-method solver (COMSOL Multiphysics) [28]. The GVD remains normal while the TOD is anomalous across the entire telecommunication L-band, with GVD = 50 fs$^2$/mm and TOD = -1160 fs$^3$/mm at 1600 nm. The normal GVD guarantees high gain for the Faraday wave and suppresses the formation of Turing patterns (Figure 1c). A high-resolution coherent swept wavelength interferometer (SWI) is implemented for characterizing the cold cavity properties of the designed microresonator [28]. The measured wavelength-dependent free spectral ranges (FSRs), overlaid with the results of simulations accounting for both the GVD and the TOD, are shown in the inset of Figure 2a. A good agreement between the experimental and simulation results is achieved. Figure 2b shows the mismatch between the signal-pump and pump-idler detuning, $\omega_N + \omega_{-N} - 2\cdot\omega_0$, as well as the FSR as a function of the azimuthal mode number. The crossing between the mismatch and the FSR denotes the region in which the first Kerr frequency comb sidebands are expected to emerge from the Faraday ripples. For the pump wavelength of 1600 nm, the numerical calculation predicts the sideband mode number N = 480, while the experimentally measured sidebands are clustered around N = 570 (Figure 2c). On the other hand, if dispersion orders higher than TOD are not considered in the numerical calculation, the obtained sideband mode number is N = 590 (red dashed line in the inset of Figure 2b) and a better agreement with the experimental sideband mode number is obtained. Thus, we attributed this discrepancy to dispersion orders higher than TOD in the numerical simulation.

The frequency of the Faraday ripples, or primary comb spacing, can be tuned by changing the pump frequency. As the pump frequency increases, the GVD decreases monotonically, resulting in a positive correlation between the primary comb spacing and the pump frequency, according to Equation 2. Figure 3 shows this correlation, obtained by pumping the microresonator's 20 different modes around 1590 nm and measuring the primary comb spacing using a high-resolution optical spectrum analyzer (OSA). Linear regression analysis on the measured data yields a good adjusted R-squared value of 99.3%. The fitted slope is 1.32 +/- 0.3, well-matching the slope of 1.27 predicted from the numerical calculation (Figure 2b). Setting the pump frequency to 189.8 THz (pump wavelength of 1580 nm) results in the primary comb spacing of 40.2 THz, with the overall bandwidth spanning 2/3 of an octave. By pumping the microresonator harder with an on-chip pump power of 30 dBm, subsequent non-degenerate FWM occurrs and three Kerr frequency combs are formed, with center wavelengths of 1304 nm, 1580 nm, and 2002 nm (Figure 4). Merging of the three Kerr frequency combs is not yet observed, but is expected to occur with a further increase in the on-chip pump power [21]. With its optical power concentrated on the two spectral edges, the merged broadband Kerr frequency comb is expected to be advantageous for reducing the complexity of the microresonator self-referencing scheme [13]. Importantly, each segment of the multispectral Kerr frequency comb covers an important telecommunication band, including the traditional O and C/L bands, as well as the emerging 2-µm channel, which is compatible with group IV photonics [30]. Both the O band and the C/L band Kerr frequency combs span more than 8 THz, while the 2 µm Kerr frequency comb spans a narrower 4 THz owing to the aberration in the near infrared (NIR) coupling optics. The comb span is defined as the bandwidth in which the comb power remains higher than the OSA noise level of -60 dBm. Furthermore, characteristics of the O band and the C/L band Kerr frequency combs are individually investigated by performing cw heterodyne beat note measurements and by measuring amplitude noise spectra with a scan range much wider than the cavity linewidth. Besides the beat note of the cw laser with the pump laser (Figure 5a), beat notes between the cw laser and different comb lines in the C/L band (Figure 5b) and the O band (Figure 5c) are also measured. All beat notes exhibit the same linewidth of 350 kHz, limited by the mutual coherence between the cw laser and the pump laser. Neither additional linewidth broadening of the comb lines relative to the pump nor multiple beat notes are observed, confirming that the comb lines exhibit a similar level of phase noise as the pump. Besides good comb line coherence, the absence of comb breathing and a sub-comb offset is also important, as they degrade the quality of telecommunication [31]. This is independently confirmed by performing amplitude noise measurements, and no peaks or excess noise are observed for frequencies reaching 2010 MHz (about ten times the cavity linewidth), for both the C/L band (inset of Figure 5b) and O band (inset of Figure 5c) Kerr frequency combs.

In summary, we examine and demonstrate the first multispectral Kerr frequency comb initiated by Faraday ripples. The momentum conservation, or phase matching condition, is fulfilled by dispersion modulation of the high-Q microresonator. Our use of a $Si_3N_4$ planar microresonator has the advantage of straightforward dispersion management by the design of waveguide geometry; the same principle can be applied to other types of microresonators by engineering the cavity structures [32–34]. Choosing a proper pump frequency, we demonstrate that the multispectral Kerr frequency comb covers important telecommunication bands, including the traditional O and C/L bands as well as the emerging 2-µm channel, which is compatible with group IV photonics. Furthermore, we confirm the comb coherence as well as the absence of comb breathing and a sub-comb offset, by performing cw heterodyne beat note and amplitude noise spectra measurements. The reported multispectral Kerr frequency comb initiated by Faraday ripples is a promising platform for high-capacity coherent telecommunication and self-referenced microresonator-based optical frequency combs.

1. C. Y. Wang, T. Herr, P. Del'Haye, A. Schliesser, J. Hofer, R. Holzwarth, T. W. Hänsch, N. Picqué, and T. J. Kippenberg, "Mid-infrared optical frequency combs at 2.5 µm based on crystalline microresonators", Nat. Commun. 4, 1345 (2013).
2. A. A. Savchenkov, V. S. Ilchenko, F. D. Teodoro, P. M. Belden, W. T. Lotshaw, A. B. Matsko, and L. Maleki, "Generation of Kerr combs centered at 4.5 µm in crystalline microresonators pumped with quantum-cascade lasers", Opt. Lett. 40, 3468 (2015).
3. A. G. Griffith, R. K. W. Lau, J. Cardenas, Y. Okawachi, A. Mohanty, R. Fain, Y. H. D. Lee, M. Yu, C. T. Phare, C. B. Poitras, A. L. Gaeta, and M. Lipson, "Silicon-chip mid-infrared frequency comb generation," Nat. Commun. 6, 6299 (2015).
4. I. S. Grudinin, K. Mansour, and N. Yu, "Properties of fluoride microresonators for mid-IR applications", Opt. Lett. 41, 2378 (2016).
5. S.-W. Huang, J. Yang, J. Lim, H. Zhou, M. Yu, D.-L. Kwong, and C. W. Wong, "A low-phase-noise 18 GHz Kerr frequency microcomb phase-locked over 65 THz", Sci. Rep. 5, 13355 (2015).
6. W. Liang, D. Eliyahu, V. S. Ilchenko, A. A. Savchenkov, A. B. Matsko, D. Seidel, and L. Maleki, "High spectral purity Kerr frequency comb radio frequency photonic oscillator", Nat. Commun. 6, 7957 (2015).
7. K. Saha, Y. Okawachi, B. Shim, J. S. Levy, R. Salem, A. R. Johnson, M. A. Foster, M. R. E. Lamont, M. Lipson, and A. L. Gaeta, "Modelocking and femtosecond pulse generation in chip-based frequency combs", Opt. Express 21, 1335 (2013).
8. X. Xue, Y. Xuan, Y. Liu, P.-H. Wang, S. Chen, J. Wang, D. E. Leaird, M. Qi, and A. M. Weiner, "Mode-locked dark pulse Kerr combs in normal-dispersion microresonators", Nat. Photon. 9, 594 (2015).
9. S.-W. Huang, H. Zhou, J. Yang, J. F. McMillan, A. Matsko, M. Yu, D.-L. Kwong, L. Maleki, and C. W. Wong, "Mode-locked ultrashort pulse generation from on-chip normal dispersion microresonators", Phys. Rev. Lett. 114, 053901 (2015).
10. X. Yi, Q.-F. Yang, K. Y. Yang, M.-G. Suh, and K. Vahala, "Soliton frequency comb at microwave rates in a high-Q silica microresonator", Optica 2, 1078 (2015).
11. V. Brasch, M. Geiselmann, T. Herr, G. Lihachev, M. H. P. Pfeiffer, M. L. Gorodetsky, and T. J. Kippenberg, "Photonic chip-based optical frequency comb using soliton Cherenkov radiation", Science 351, 357 (2016).
12. S.-W. Huang, J. Yang, M. Yu, B. H. McGuyer, D.-L. Kwong, T. Zelevinsky, and C. W. Wong, "A broadband chip-scale optical frequency synthesizer at 2.7×10-16 relative uncertainty", Sci. Adv. 2, e1501489 (2016).
13. P. Del'Haye, A. Coillet, T. Fortier, K. Beha, D. C. Cole, K. Y. Yang, H. Lee, K. J. Vahala, S. B. Papp, and S. A. Diddams, "Phase-coherent microwave-to-optical link with a self-referenced microcomb", Nat. Photon. 10, 516 (2016).
14. J. Pfeifle, A. Coillet, R. Henriet, K. Saleh, P. Schindler, C. Weimann, W. Freude, I. V. Balakireva, L. Larger, C. Koos, and Y. K. Chembo, "Optimally coherent Kerr combs generated with crystalline whispering gallery mode resonators for ultrahigh capacity fiber communications", Phys. Rev. Lett. 114, 093902 (2015).
15. C. Bao, P. Liao, A. Kordts, M. Karpov, M. H. P. Pfeiffer, L. Zhang, Y. Yan, G. Xie, Y. Cao, A. Almaiman, M. Ziyadi, L. Li, Z. Zhao, A. Mohajerin-Ariaei, S. R. Wilkinson, M. Tur, M. M. Fejer, T. J. Kippenberg, A. E. Willner, "Demonstration of optical multicasting using Kerr frequency comb lines", Opt. Lett. 41, 3876 (2016).
16. Y. Okawachi, M. R. E. Lamont, K. Luke, D. O. Carvalho, M. Yu, M. Lipson, and A. L. Gaeta, "Bandwidth shaping of microresonator-based frequency combs via dispersion engineering", Opt. Lett. 39, 3535 (2014).
17. A. B. Matsko, A. A. Savchenkov, S.-W. Huang, and L. Maleki, "Clustered frequency comb", Opt. Lett., accepted.

18. A. Coillet, I. Balakireva, R. Henriet, K. Saleh, L. Larger, J. M. Dudley, C. R. Menyuk, and Y. K. Chembo, "Azimuthal Turing patterns, bright and dark cavity solitons in Kerr combs generated with whispering-gallery-mode resonators", IEEE Photon. J. 5, 6100409 (2013).
19. S.-W. Huang, J. Yang, S.-H. Yang, M. Yu, D.-L. Kwong, T. Zelevinsky, M. Jarrahi, and C. W. Wong, "Universally stable microresonator Turing pattern formation for coherent high-power THz radiation on-chip", arXiv:1603.00948 (2016).
20. A. Dutt, K. Luke, S. Manipatruni, A. L. Gaeta, P. Nussenzveig, and M. Lipson, "On-chip optical squeezing", Phys. Rev. Applied 3, 044005 (2015).
21. T. Herr, K. Hartinger, J. Riemensberger, C. Y. Wang, E. Gavartin, R. Holzwarth, M. L. Gorodetsky, and T. J. Kippenberg, "Universal formation dynamics and noise of Kerr-frequency combs in microresonators", Nat. Photon. 6, 480 (2012).
22. J. A. Armstrong, N. Bloembergen, J. Ducuing, and P. S. Pershan, "Interaction between light waves in a nonlinear dielectric", Phys. Rev. 127, 1918 (1962).
23. M. Faraday, "On a peculiar class of acoustical figures; and on certain forms assumed by groups of particles upon vibrating elastic surfaces", Philos. Trans. R. Soc. Lond. 121, 299 (1831).
24. P. Engels, C. Atherton, and M. A. Hoefer, "Observation of Faraday waves in a Bose-Einstein condensate", Phys. Rev. Lett. 98, 095301 (2007).
25. M. Conforti, A. Mussot, A. Kudlinski, and S. Trillo, "Modulational instability in dispersion oscillating fiber ring cavities", Opt. Lett. 39, 4200 (2014).
26. N. Tarasov, A. M. Perego, D. V. Churkin, K. Staliunas, and S. K. Turitsyn, "Mode-locking via dissipative Faraday instability" Nat. Commun. 7, 12441 (2016).
27. F. Copie, M. Conforti, A. Kudlinsky, A. Mussot, and S. Trillo, "Competing Turing and Faraday instabilities in longitudinally modulated passive resonators", Phys. Rev. Lett. 116, 143901 (2016).
28. S.-W. Huang, H. Liu, J. Yang, M. Yu, D.-L. Kwong, and C. W. Wong, "Smooth and flat phase-locked Kerr frequency comb generation by higher order mode suppression", Sci. Rep. 6, 26255 (2016).
29. S. Coen, and M. Haelterman, "Modulational instability induced by cavity boundary conditions in a normally dispersive optical fiber", Phys. Rev. Lett. 79, 4139 (1997).
30. R. Soref, "Group IV photonics: Enabling 2 µm communications", Nat. Photon. 9, 358 (2015).
31. C. Bao, P. Liao, L. Zhang, Y. Yan, Y. Cao, G. Xie, A. Mohajerin-Ariaei, L. Li, M. Ziyadi, A. Almaiman, L. C. Kimerling, J. Michel, and A. E. Willner, "Effect of a breather soliton in Kerr frequency combs on optical communication systems", Opt. Lett. 41, 1764 (2016).
32. L. Zhang, A. M. Agarwal, L. C. Kimerling, and J. Michel, "Nonlinear group IV photonics based on silicon and germanium: from near-infrared to mid-infrared", Nanophotonics 3, 247 (2014).
33. I. S. Grudinin, and N. Yu, "Dispersion engineering of crystalline resonators via microstructuring", Optica 2, 221 (2015).
34. K. Y. Yang, K. Beha, D. C. Cole, X. Yi, P. Del'Haye, H. Lee, J. Li, D. Y. Oh, S. A. Diddams, S. B. Papp, and K. J. Vahala, "Broadband dispersion-engineered microresonator on a chip", Nat. Photon. 10, 316 (2016).

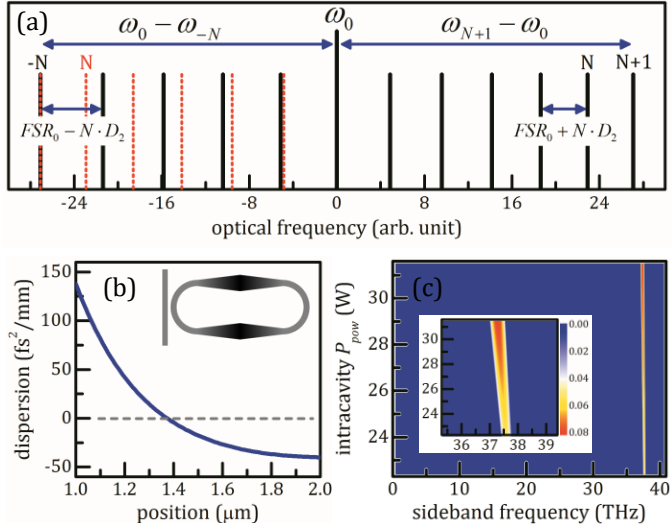

**Fig. 1.** (a) Principle of the Kerr frequency comb generation, assuming that the azimuthal mode numbers of the signal and idler are not symmetrically located with respect to the pump mode. Owing to the cavity GVD, the mismatch between the signal-pump and pump-idler detuning increases as a quadratic function of the azimuthal mode number, eventually approaching the FSR of the (N+1)$^{th}$ signal mode at $N = \left[\sqrt{\frac{FSR_0}{-D_2}} - 1\right]$. Consequently, the energy conservation law of $\omega_{N+1} + \omega_{-N} - 2\cdot\omega_0 \approx 0$ is satisfied. However, the asymmetry in the signal and idler mode numbers renders the momentum conservation law violated unless an azimuthal modulation of the cavity parameters is introduced to provide an additional wavevector for the QPM. The non-equidistance of the modes is $D_2 = -\frac{\beta_2 c FSR_0^2}{n}$. (b) COMSOL-modeled GVD of the $Si_3N_4$ waveguide with respect to the waveguide width, taking into consideration both the waveguide dimensions and the material dispersion. With the waveguide width varying from 1 μm to 2 μm, the simulated GVD changes from normal at 140 fs$^2$/mm to anomalous at -40 fs$^2$/mm. Inset: Schematic of our dispersion-modulated single-mode $Si_3N_4$ microresonator. (c) Analytic OPO gain as a function of the sideband frequency and intra-cavity pump power. Inset: Zoom-in map of the roundtrip gain coefficient, showing the gain peak of Faraday ripples at ~37.5 THz.

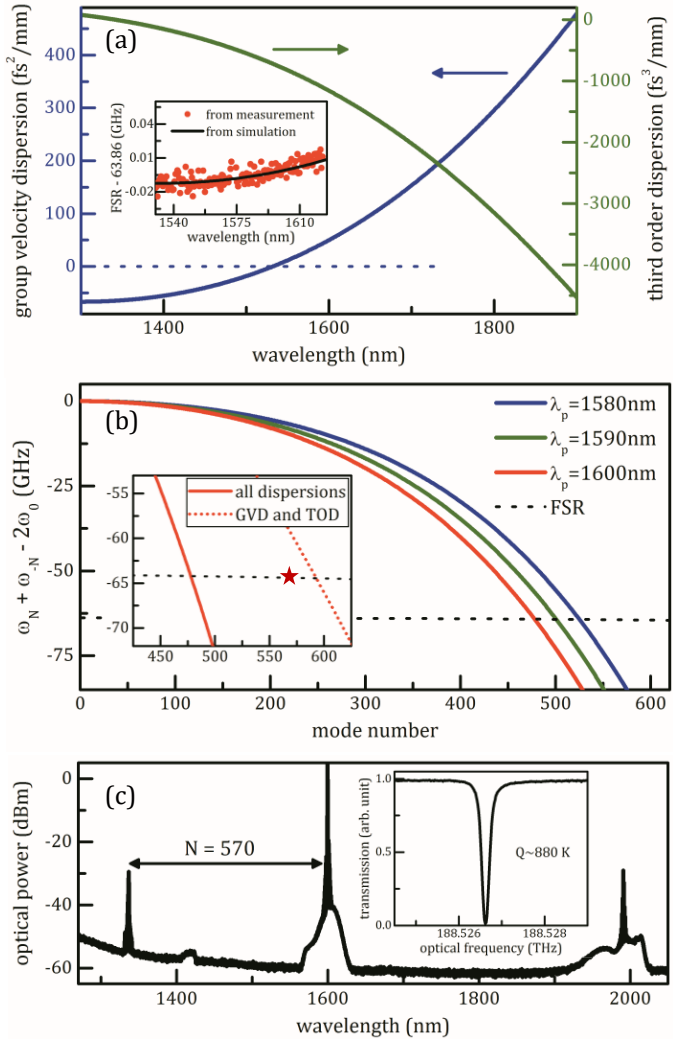

**Fig. 2.** (a) Path-averaged GVD and TOD of our dispersion-modulated single-mode microresonator, showing a monotonic GVD change from 20 fs$^2$/mm at 1560 nm to 50 fs$^2$/mm at 1600 nm. Inset: The measured wavelength-dependent FSRs (red dots), showing a good agreement with the simulation results (black line) taking into account both the GVD and the TOD. (b) The mismatch between the signal-pump and pump-idler detuning and the FSR, as a function of the azimuthal mode number. The crossing between the mismatch and the FSR denotes the region in which the first Kerr frequency comb sidebands are expected to emerge from the Faraday ripples. Inset: A magnified view of the crossing points, showing the experimentally measured point (star) between the values numerically obtained with (solid line) and without (dashed line) higher-order dispersions. (c) Example optical spectrum of Faraday ripples pumped at 1600 nm, showing the first pair of sidebands at the azimuthal mode N = 570. Inset: Cold cavity resonance at 1600 nm, measuring a loaded Q of 880,000 with critical coupling.

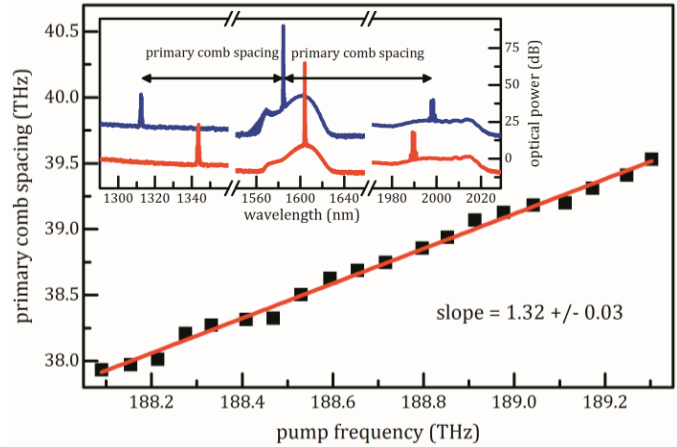

**Fig. 3.** Frequency of the Faraday ripples, or primary comb spacing, linearly related to the pump frequency. The fitted slope is 1.32 +/- 0.3, well-matching the slope of 1.27 predicted from the numerical calculation. Inset: Example optical spectra of the Faraday ripples pumped at 1585 nm (blue line) and 1605 nm (red line).

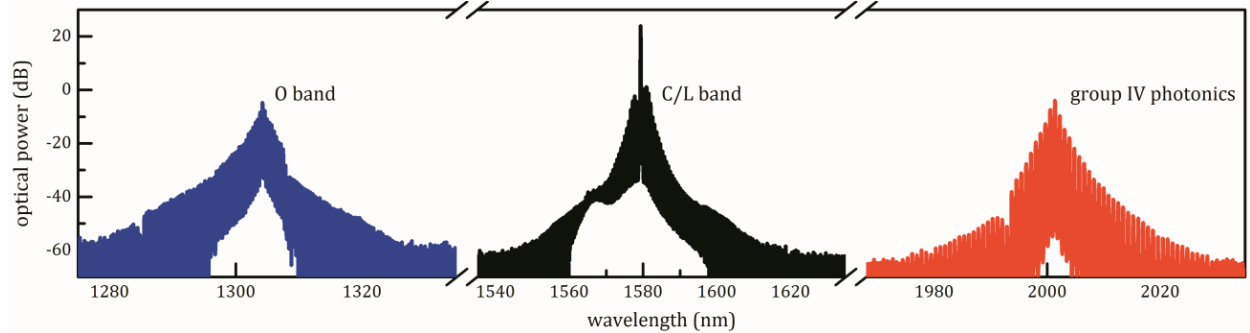

**Fig. 4.** Broadband multispectral Kerr frequency comb, with each segment covering an important telecommunication band, including the traditional O and C/L bands as well as the emerging 2-μm channel, which is compatible with group IV photonics. Both the O band and the C/L band Kerr frequency combs span more than 8 THz above the OSA noise level of -60 dBm, while the 2 μm Kerr frequency comb spans a narrower 4 THz owing to the aberration in the NIR coupling optics.

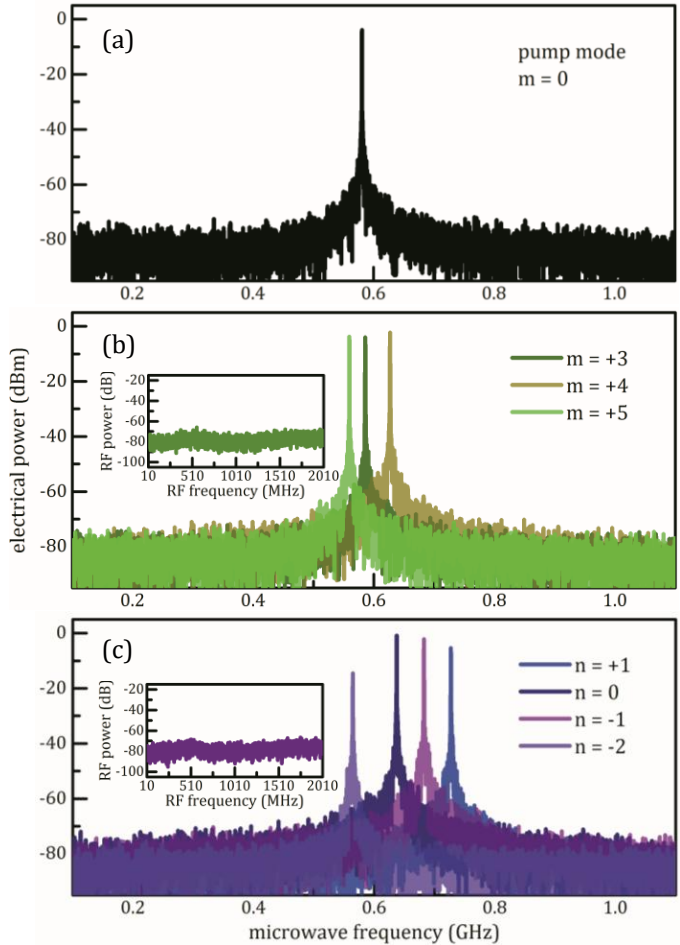

**Fig. 5.** Good coherence and absence of comb breathing and sub-comb offset are confirmed by the cw heterodyne beat note and amplitude noise spectra measurements. (a) Heterodyne beat note between the pump laser and an independent narrow linewidth cw laser, measuring a linewidth of 350 kHz. (b) Heterodyne beat notes between the cw laser and three other C/L band Kerr frequency comb lines, all measuring the same 350 kHz linewidth which is limited by the mutual coherence between the cw laser and the pump laser. Inset: RF amplitude noise of the C/L band Kerr frequency comb, showing no peaks and excess noise up to 2010 MHz. (c) Heterodyne beat notes between the cw laser and four different O band Kerr frequency comb lines, all measuring the same 350 kHz linewidth which is limited by the mutual coherence between the cw laser and the pump laser. Inset: RF amplitude noise of the O band Kerr frequency comb, showing no peaks and excess noise up to 2010 MHz. $n = 0$ corresponds to the comb line at the center wavelength of 1304 nm.